# Three-dimensional patterning of solid microstructures through laser reduction of colloidal graphene oxide in liquid-crystalline dispersions

Bohdan Senyuk[1,2], Natnael Behabtu[2], Angel Martinez[1], Taewoo Lee[1], Dmitri E. Tsentalovich[2], Gabriel Ceriotti[3], James M. Tour[3], Matteo Pasquali[2,3] and Ivan I. Smalyukh[1,4,5,*]

[1] *Department of Physics, University of Colorado, Boulder, Colorado 80309, USA*

[2] *Department of Chemical and Biomolecular Engineering, Rice University, Houston, Texas 77005, USA*

[3] *Department of Chemistry and the R. E. Smalley Institute for Nanoscale Science and Technology, Rice University, Houston, Texas 77005, USA*

[4] *Department of Electrical, Computer, and Energy Engineering, Materials Science and Engineering Program, and Liquid Crystal Materials Research Center, University of Colorado, Boulder, CO 80309, USA*

[5] *Renewable and Sustainable Energy Institute, National Renewable Energy Laboratory and University of Colorado, Boulder, Colorado 80309, USA*

*e-mail: ivan.smalyukh@colorado.edu

## Abstract

Graphene materials and structures have become an essential part of modern electronics and photovoltaics. However, despite many production methods, applications of graphene-based structures are hindered by high costs, lack of scalability and limitations in spatial patterning. Here we fabricate three-dimensional functional solid microstructures of reduced graphene oxide in a lyotropic nematic liquid crystal of graphene oxide flakes using a pulsed near-infrared laser. This reliable, scalable approach is mask-free, does not require special chemical reduction agents, and can be implemented at ambient conditions starting from aqueous graphene oxide flakes. Orientational ordering of graphene oxide flakes in self-assembled liquid-crystalline phases enables laser patterning of complex, three-dimensional reduced graphene oxide structures and colloidal particles, such as trefoil knots, with "frozen" orientational order of flakes. These structures and particles are mechanically rigid and range from hundreds of nanometres to millimetres in size, as needed for applications in colloids, electronics, photonics and display technology.

## Introduction

Owing to unique physical properties[1,2], graphene-based materials and devices are shaping the future of photonics, electronics, optoelectronics and energy applications[2-8]. For example, they promise to replace rare and brittle indium tin oxide transparent electrodes in the display industry[1,3]. The most effective, low-cost and scalable approach to prepare graphene based materials is the reduction of graphene oxide (GO)[9,10], which can be produced in abundance. Graphene oxide can be reduced by multiple methods[5-26], including chemical reduction[10], photoreduction[6-8,10-14] and even reduction mediated by biological microorganisms[15]. Most of these methods require specific precursors and conditions, which make them expensive and unsuitable for large-scale production. Multiple lithographic methods[5,16] have been developed for post-production of reduced graphene oxide (rGO) and its micropatterning into useful structures and assemblies, but they are time and resource consuming and limited by advances in mask production. Recently, reduction of GO and micropatterning was achieved using laser irradiation[6-8,10-13,17-23,27]. However, its use so far has been limited to reduction and direct writing of patterns in dry two-dimensional (2D) solid films of GO.

      Here we produce fully three-dimensional (3D) functional solid microstructures of rGO in an aqueous nematic liquid crystal (LC) of 2D GO flakes using pulsed near-infrared laser scanning and characterize them using nonlinear optical microscopy and other techniques. We show that photoluminescence of laser-reduced rGO is increased compared with pristine GO, which allows for using the same excitation beam for simultaneous reduction of GO into desired microstructures and their subsequent nonlinear photoluminescence imaging. Orientational order of LC phases promotes homogeneous internal structure of complex 3D rGO patterns. Our approach is reliable and scalable, does not require special conditions or chemical reducing agents and can be used in ambient conditions in high concentration aqueous dispersions of GO flakes. Because laser reduction with tightly focused femtosecond light is an intrinsically depth-resolved nonlinear process, it can produce layered, sandwiched, as well as fully 3D microstructures, including nonplanar surfaces such as those with the topology of trefoil torus knots, by programmable spatial steering of a tightly focused laser beam, without lithographic masks. Various 3D topologically nontrivial solid microstructures of rGO obtained with this approach can be used in electronic and optoelectronic devices and in basic research.

## Results

**Laser reduction of colloidal graphene oxide.** Solid microstructures of rGO were produced in concentrated aqueous dispersions of pristine GO flakes[28,29] (see Methods) sandwiched in between two clean, untreated glass substrates (Fig. 1a). Graphene oxide flakes (Fig. 2a) are hundreds of nanometres wide graphene sheets with a basal plane and edges decorated by oxygen-containing groups including carboxyls, hydroxyls, epoxides and others[10], which cause screened electrostatic repulsion that stabilize the colloidal flakes against aggregation prompted by interlayer forces between flakes, making them both strongly hydrophilic and stable in colloidal dispersions. They can be dispersed in deionized water to form different lyotropic discotic LC phases at higher concentrations[29–32]. In this work, we use aqueous GO flakes at 0.25-1 wt% to achieve a stable self-assembled nematic LC phase (Fig. 1b,c and Supplementary Fig. 1). Flat GO flakes align edge-on at the interface with glass substrates[29,31] and spontaneously orient in the bulk so that normals to the surfaces of disk-like planes point roughly along the LC director **n**, which describes the average local orientation of normals to the flakes[29]. Nematic ordering of GO flakes (Supplementary Fig. 1) is important because it promotes homogeneous internal structure in produced rGO micropatterns (Fig. 1b,c). Nematic samples of thickness within $d \approx 10\text{-}30$ μm have a yellow brown colour and are partially transparent at wavelengths ranging from about 400 nm to far in the near infrared region (bright areas in Fig. 1b and its inset), showing maximum absorbance at about 240 nm (Supplementary Fig. 2a) due to $\pi\text{-}\pi^*$ electron transitions of carbon-carbon bonds[10]. Figure 1b shows a transmission-mode bright-field microscopy texture of a GO nematic dispersion when illuminated by the femtosecond laser light at a wavelength 850 nm (Supplementary Fig. 2a) and the inset shows the same texture when observed under white light. Broad-band nonlinear photoluminescence[29] is detected from samples (Fig. 2e) under illumination by a femtosecond pulsed laser beam of low power (Methods and Supplementary Fig. 3) at 850 nm, which allows for nonlinear, intrinsically depth-resolved imaging of nematic textures[33] in GO LCs[29]. On the other hand, illumination of a GO nematic with a pulsed laser beam of high laser energy density (fluence) results in a permanent physical change with properties drastically different from the pristine samples. We attribute this permanent change of illuminated area to a local reduction of GO flakes by a laser beam[6–8,10–13,27].

Figure 1b shows a bright-field micrograph of a square area illuminated by a laser beam of high fluence, $E \approx 60\text{-}70$ mJ cm$^{-2}$. Its appearance is dark (Fig. 1b) and its transmittance $T$ is lower (Figs 1d, 3a) than through the unaffected bright area. At the same time, the laser-treated area produces significantly higher intensity of photoluminescence, $I_{PL}$ (Figs 1e, 2e, 3c), allowing for the simultaneous nonlinear photoluminescence imaging of a produced pattern (Fig. 1c). Owing to the intrinsic depth-resolved nature of this nonlinear optical process[34,35], the laser reduction of aqueous GO is spatially localized in a small volume of about 300 nm in diameter defined by optical resolution[33–35] and called a voxel. This 3D localization allows for voxel-by-voxel micro-patterning of rGO not only in the plane of the sample (Fig. 1b,c,f) but also across the sample thickness (Fig. 1g), which sets our approach apart from previous techniques[6–8,10–13,17–23,27] of photothermal or photo-induced reduction of GO and 2D micropatterning in thin dried films. Three planes of rGO, whose thickness is determined by optical resolution, were produced at different heights within a thick GO nematic dispersion sample (Fig. 1f) using a tightly focused laser beam. This is achieved without employing multi-layer lithography[5]. Programmable control of a scanning laser beam allows for precise voxel-by-voxel reduction of GO and micropatterning not only in-plane at different heights (Fig. 1f,e) but also continuous reduction across a thick layer of a GO nematic (Supplementary Fig. 4).

**Characterization of rGO.** To probe how a femtosecond pulsed laser beam at 850 nm modifies aqueous GO samples, we use ultraviolet-visible (Fig. 2e and Supplementary Fig. 2), Raman (Fig. 2f and Supplementary Fig. 5) and X-ray photoelectron (XPS; Fig. 2g,h) spectroscopies, as well as optical (Fig. 3a,b) and nonlinear photoluminescence (Fig. 3c,d) imaging (Methods). Raman spectra (Fig. 2f) collected from an untreated GO area and an rGO region (Fig. 1b) show pronounced peaks located at about 1,357 cm$^{-1}$ and 1,598 cm$^{-1}$ corresponding to a D band induced by structural disorder and a G band associated with the vibration of $sp^2$-bonded carbon atoms[8,10]. The background fluorescence in a Raman signal is significantly quenched in the rGO areas (Supplementary Fig. 5), depending on the duration of irradiation or dwell time at constant fluence. There is no significant shift of D and G peaks or change in the ratio between their intensities, but the absolute intensity of the Raman signal is increased in the rGO regions (Fig. 2f). Compared with the GO areas, rGO regions show sharper D and G bands (Fig. 2f), which can be attributed to removal of oxygen-containing groups[8,10], as we additionally verify using XPS. Figure 2g,h shows XPS spectra, which clearly indicate the removal of oxygen-containing groups during the 850-nm laser irradiation of GO in water. XPS survey spectra (Fig. 2g) show signals of carbon and oxygen from both GO and rGO, however, the intensity of the oxygen peak O1s from an rGO area is significantly decreased, indicating that oxygen was partially removed. The high-resolution C1s XPS spectra (Fig. 2h) were deconvoluted into peaks corresponding to carbon in, respectively, C-C, C-OH, C=O and O=C-OH functional groups, which allows us to estimate the relative content of carbon not bound to oxygen as ~ 45.8% in GO and ~ 58% in the rGO area. The observed levels of reduction (Fig. 2g,h) are similar to those reported in previous studies of 2D laser reduction in solid thin films[11]. In addition, measurements of resistance (Supplementary Fig. 6) of produced plain rGO stripe microstructures, which were performed using methodology similar to that developed for the study of photoreduction of GO in solid thin films[11], indicate that rGO microstructures are more conductive as compared with the electrically insulating GO.

Differences in physical properties between pristine and laser-reduced GO are also detected by optical spectroscopy (Fig. 2e) and optical bright-field and nonlinear photoluminescence microscopy (Fig. 3). The absorbance of rGO areas increases in the visible range and the onset of the absorbance peak is redshifted (Supplementary Fig. 2b). The transmittance of rGO within the visible range is decreased (Fig. 3a) as they change appearance from transparent yellowish to opaque dark brown (Fig. 3b), controlled by the dwell time of the scanning beam. At the same time, the intensity of nonlinear photoluminescence from rGO areas increases (Figs 2e, 3c,d), with its maximum moving to shorter wavelengths[36], depending on the dwell time. The increase of photoluminescence from laser-reduced areas was also observed in dried films of giant[31] GO flakes (Supplementary Fig. 7); GO can be laser-reduced even within just a single giant GO flake (Supplementary Fig. 7a). Interestingly, the transmittance, $T$, of rGO beyond a threshold irradiance continuously decreases almost linearly with irradiation time to a minimum (Fig. 3a,b), but the intensity of photoluminescence increases with irradiation time and reaches saturation after a rather short dwell time (Fig. 3c). This behaviour of $I_{PL}$ may provide insights into the physical origins of photoluminescence from graphene-based materials[29,37] and can be related to increasing the fraction of $sp^2$-hybridized carbon during reduction, which increases both absorbance and photoluminescence intensity[36,38]. Optical characterization reveals a qualitative picture of the reduction process and formation of solid structures of rGO flakes. Pristine GO flakes (Fig. 2a) in the lyotropic nematic LC are oriented, on average, in the same direction and are kept apart by screened electrostatic repulsions between the charged GO surface, with the average spacing between the hydrophilic flakes determined by the volume fraction of GO in the dispersion (Fig. 2c). Laser-induced reduction largely removes oxygen-containing groups from the GO flakes (Fig. 2b,g,h) making them more hydrophobic and weakening the electrostatic repulsions (Fig. 2d). Consequently, hydrophobic rGO flakes overlap with each other and form a solid continuous structure (Fig. 2d and Supplementary Fig. 8c) defined by the 3D illumination pattern. Dependence of photoluminescence from rGO as well as white-light transmission on the irradiation time (Fig. 3) and microscopic observations (Fig. 1b,c) allow concluding that for the used parameters of the irradiation process the onset of reduction is not very sharp. Experiments also show that at high laser powers, $E \gg 100$ mJ cm$^{-2}$, the process most probably can become more violent with rapid expulsion of water and oxygen and damaging integrity of a sample. We therefore constrain the irradiation parameters so that the onset of reduction and rGO flake aggregation into microstructures is smooth and continuous and dependent on duration of irradiation (Fig. 3). Importantly, the laser patterning is done through scanning of a tightly focused beam to reduce the structure voxel-by-voxel, without causing bubbles or disrupting orientational ordering of the flakes. Progressive reduction of GO flakes by the scanned beam focused to a volume <1 $\mu m^3$ and diffusion of flakes around the region of scanning allow us to avoid flows and non-homogeneity of the GO dispersion in the region of reduction. As the GO flakes within the micro-patterned structure collapse atop of each other due to lowering of electrostatic repulsion forces caused by reduction, water is displaced from inter-flake regions continuously during the scanning, so that the effect of these localized flows on GO flake ordering can be neglected. Solid structures of rGO are rigid and preserve their shape even under mechanical stress induced, for example, by the flow of LC during shear of confining glass plates (Supplementary Fig. 8a,b). The rGO structures produced in the bulk of the LC fluid are permanent colloidal objects and, at normal conditions, do not "dissolve" in an aqueous GO dispersion. For example, such microstructures in the form of square-shaped sheets of rGO could be re-dispersed in water and then transferred onto solid substrates for scanning electron microscopy (SEM) imaging, an example of which is shown in the inset of Fig. 1c. Thus, this approach may allow for facile production of not only surface-attached micro- and nano-structures, but also complex-shaped colloids[34] made of ribbons or tubes of rGO that could be then immersed into different fluid host media. SEM imaging of solid and rigid rGO microstructures (Supplementary Fig. 8c) retrieved from a sample after reduction show a fairly homogeneous surface morphology (the inset of Fig. 1c and Supplementary Fig. 9). The folded wrinkles and ordering of overlapped rGO flakes visible in the SEM images of the solid structures resemble smooth changes of orientation due to their on-average alignment with normal along the discotic nematic director **n** in the original fluid nematic of GO, which was "frozen" into a solid structure after flakes collapsed atop of each other due to weakening of electrostatic repulsions caused by laser reduction. Absence of visible deep cracks or pores suggests that the expulsion of water and oxygen during the laser-induced reduction, at discussed parameters, was not a rapid or violent process and did not introduce additional disorder or defects in the nematic molecular alignment field. Our findings indicate that, by leveraging the intrinsic ordering of GO flakes in a nematic phase, the shape anisotropy of ribbon-like cross-sections and the degree of reduction may potentially be controlled through varying the GO flakes concentration beyond the critical value of nematic formation.

**Microstructures of rGO.** Reduction of GO flakes in aqueous discotic nematic samples in a spatially confined nanosized spot allows for direct and precise micropatterning of various 3D solid structures of rGO (Figs 4, 5 and Supplementary Figs 4, 10) using computer-directed voxel-by-voxel scanning of a laser beam. The degree of reduction can be controlled by changing the fluence and irradiation/dwell time (Fig. 3 and Supplementary Fig. 10). Figure 4 shows different complex multi-level patterns of rGO produced in a bulk and at a surface of aqueous GO nematic and imaged by optical and photoluminescence microscopy. Using transmission-mode bright-field microscopy, one can see all the elements of micropatterns at the same time, but blurred depending on the location of the microscope's focal plane relative to the features of micropatterns (Fig. 4a,d,f): an image is focused at a single level (Fig. 4a) of rGO "nanostairs" going from the top to the bottom substrate, or it is focused on the different parts of the text (Fig. 4d,f) micropatterned on different levels in a GO nematic cell. At the same time, confocal photoluminescence microscopy allows the detection of the photoluminescence signal only from specific planes at different depths across the multi-level structures of rGO (see, for example, Fig. 4b,c,e,g), while other details outside of the laser scanned area are invisible (Fig. 4e,g and Supplementary Movie 1). This property of selective reduction can be used in high-capacitance all carbon[6] information storage devices. Figure 4h–j shows an example structure that can be used as prototypes of thin-film 2D or 3D resistor elements. Intersecting multi-level or multi-layer structures (Fig. 4k–o and Supplementary Movie 2), where insulating GO is sandwiched in between the more conductive rGO layers (Supplementary Fig. 6), can be envisaged as prototypes of electronic elements for applications in electronics and optoelectronics, such as thin film capacitors[6,18,20] and field transistors[6]. Importantly, our approach is suitable for direct mask-free production of not only in-plane or sandwiched structures and devices[6,7,12,14,22] but also fully 3D interconnecting patterns, structures and nonplanar surfaces, which is one of the key advances as compared with, for example, in-plane structures presented in a pioneering study on photoreduction of GO in solid thin films[11]. To further demonstrate this capability, we produce an array of rGO trefoil knots (Fig. 5 and Supplementary Movie 3), which have topologically nontrivial 3D shapes described by a set of parametric equations as $\mathbf{r} = L[2.1(\cos \phi - 2.25 \cos 2\phi), 2.1(\sin \phi + 2.25 \sin 2\phi), 6 \sin 3\phi]$, where $L$ determines the outer size of the colloidal knot and the parameter $\phi \in [0, 2\pi]$[34]. Bright-field optical (Fig. 5a–c) and photoluminescence (Fig. 5e–g) images of the rGO trefoil knot are obtained when focusing at the top, middle and bottom of the knot, respectively; a 3D perspective view of the knot (Fig. 5h and Supplementary Movie 3) was reconstructed from experimental 3D data of photoluminescence imaging. Because the internal orientation of rGO flakes within the colloidal structures matches that of the surrounding GO flakes, the fabricated knotted particles differ from the polymerized ones studied recently[34] in that they do not induce noticeable director distortions or topological defects (Fig. 5). This demonstrates that topological defects can be avoided when the easy axis (boundary conditions) for the director orientation on the surface of complex-shaped particles is such that it mostly matches that of the surrounding LC, even when the colloidal inclusions exhibit nontrivial surface topology of torus knots.

**Discussion**

We have demonstrated selective reduction of GO flakes in the bulk of an aqueous LC dispersion of GO flakes at ambient conditions using nonlinear excitation by a near-infrared femtosecond laser light, which allows for mask-free micropatterning of 3D solid planar and nonplanar structures and surfaces of reduced GO. Besides its simplicity, this reliable, scalable method does not require special conditions or chemical reduction agents and can be implemented at ambient conditions in aqueous dispersions of GO flakes. Advantages of micropatterning in a discotic nematic LC phase include the possibility of producing patterns having more homogeneous internal structure and the potential to attain "frozen" orientational order of GO flakes, which can also be modified before reduction by applying an external fields[32] or using other stimuli. Our approach can be used for scalable production of graphene-based devices for photonics, electronics[20], optoelectronics[12], information[22] and energy storage[7] applications. Furthermore, solid colloidal particles of rGO produced here are enabling new explorations in other research areas such as, for example, for probing the interplay of topologies of surfaces, molecular fields and defects in soft condensed matter[34]. In this regard, the ribbon-like shape of thin rGO colloidal structures may allow for obtaining colloids with surfaces lacking orientability, such as Möbius strips[39], which would be otherwise difficult to produce using techniques like two-photon polymerization[34].

**Methods**

**Preparation of aqueous dispersions of graphene oxide flakes.** The improved, mostly single-layer GO flakes used in this work were synthesized by methods[28] modified for large-scale production and obtained as a dispersion in deionized (DI) water at a concentration of 2.5 g l$^{-1}$ (0.25 wt%). The synthesis of GO flakes and their purification procedure was as following: 3 g of xGnP 5-µm-graphite-nanoplatelets (XG Sciences, Inc.) and 200 ml of 9:1 $H_2SO_4/H_3PO_4$ solution were added to a 1-l Pyrex beaker. The mixture was set to stir continuously using an IKA Werke RW 16 mechanical stirrer. Slowly, 9 g of $KMnO_4$ were stirred into the mixture, which turned a deep dark green colour. Precautions were taken to avoid exceeding ~ 5 wt% of $KMnO_4$ per addition and to assure that the change in colour from green to purple is complete before introducing more oxidant (this is important as solutions with more than 7 wt% of $KMnO_4$ added to $H_2SO_4$ could explode upon heating). The mouth of the beaker was covered as much as possible using aluminum foil and the reaction was heated to 40°C using a resistance-heated water bath. After 6 h, the mixture had thickened and turned from dark green to purple-pink. At this point, additional 9 g of $KMnO_4$ were added and allowed to react for another 6 h. When the reaction was completed, the resulting slurry was quenched over 200 ml of ice-water containing 5 ml of 30% $H_2O_2$. The resulting bright yellow GO suspension was purified by a series of centrifugations and resuspensions in a pure solvent. All centrifugations were performed at 4,000 r.p.m. for 90 min, and all re-suspensions were performed by shaking the centrifuged GO for ~ 4 h at 200 r.p.m. on a platform shaker using 200 ml of solvent. The product was centrifuged once as is, and centrifuged washed once in DI $H_2O$, once in 30% HCl, and twice more in DI $H_2O$. To determine the concentration of the final solution, a 10-ml aliquot was filter-washed with 200 ml of three different solvents: methanol, acetone and diethyl ether. The final material was vacuum dried at 4 Torr and room temperature overnight and weighed (25 mg).

**Sample preparation.** The aqueous dispersion of GO flakes was additionally tip-sonicated for 2 h at ~ 35 W of ultrasonic power using a Branson 250 sonifier (VWR Scientific) operating at 20 kHz and equipped with a microtip of diameter 4.8 mm, which allowed for achieving monodisperse flakes of smaller size[29]. A homogeneous nematic LC phase (Supplementary Fig. 1) was obtained at higher concentrations of GO flakes in dispersions by controlled removal of excess water from dispersions subjected to centrifugation in a Sorvall Legend 14 centrifuge (Thermo Scientific) at 12,000 r.p.m. for ~ 10-15 min. LC dispersions of GO were sandwiched between two glass substrates. Clean untreated glass substrates were used in samples for laser-induced reduction of GO flakes and optical and nonlinear microscopy observations. A gap between substrates was set using Mylar films (DuPont Teijin) of thickness 10-30 μm. Dispersions of GO flakes were sonicated in a Cole-Parmer 8891 ultrasonic bath for ~ 5 min before assembling a cell. Evaporation of water during experiments was prevented by sealing samples with a ultraviolet curable glue (NOA 63, Norland Products, Inc.). To retrieve a produced rGO microstructure for SEM imaging, a sample was disassembled and the GO dispersion, together with the rGO microstructure, was released into a Petri dish with DI water. Then, after "washing" it in DI water, the rGO microstructure with some water was soaked into a pipet tip and placed onto a silicon wafer (Supplementary Fig. 8c). Once the residual water evaporated, we performed SEM imaging.

**Laser-induced reduction and optical characterization.** The integrated 3D laser induced reduction and nonlinear photoluminescence imaging of GO aqueous samples was performed at room temperature using a multimodal nonlinear optical microscopy setup[33] coupled to an inverted Olympus microscope IX-81 (Supplementary Fig. 3). A tunable (680-1,080 nm) Ti:sapphire oscillator (140 fs, 80 MHz, Chameleon Ultra II, Coherent) was used as an excitation source. The excitation beam was directed to the sample by a system of mirrors (DMs) and focused into the sample using the Olympus high numerical aperture (NA) oil objective UPLSAPO 100×/NA = 1.4 (OL1). The spatial 3D position of the excitation beam in the volume of the sample was controlled with the galvanomirror scanning unit (Fluoview FV300, Olympus). An average laser power for photoluminescence imaging was controlled by a pair of Glan polarizer and half-wave plate to be <1 mW in a sample to prevent the photo- and thermal damage. The excitation of GO flakes was performed at 850 nm (Supplementary Fig. 2a), and the unpolarized photoluminescence light was detected in a range of ~ 400-700 nm (Fig. 2e) in a backward mode with a Hamamatsu photomultiplier tube H5784-20 (PMT1). The transmission-mode single wavelength bright-field image was collected in a forward mode with PMT2. The polarization of excitation could be varied using a half- or quarter-wave retardation plate mounted immediately before an objective. The same setup was utilized for a 3D reduction of aqueous GO flakes using a pulsed laser beam at 850 nm and laser fluence $E < 100$ mJ cm$^{-2}$. The dwell time, or scanning speed, was also controlled by a scanning unit FV300. Olympus Fluoview software was used for data acquisition and image reconstruction, and ImageJ software was used for data processing and analysis. Polarizing microscopy observations of samples in visible light and measurements of photoluminescence spectra were performed using the same IX-81 equipped with crossed polarizers, CCD camera (Flea, PointGrey) and a spectrometer USB2000 (OceanOptics) mounted onto a microscope. Raman spectra were measured using Renishaw inVia Raman microscope with excitation at 514.5 nm and low power to reduce unwanted heating or optical effects induced by a laser. XPS spectra were collected using a PHI Quantera XPS microprobe (Physical Electronics, Inc.) and SEM imaging was performed using a Zeiss Auriga FIB-SEM instrument.

**Acknowledgements**

We thank F. Mirri, Q. Liu, C. Twombly, R. Trivedi, P. Ackerman, B. Dan, A. Lee, F. Vitale and C. Young for useful discussions. We are grateful to B. Chen (Rice University Shared Equipment Authority) and F. Mirri for kind assistance with XPS measurements. We also thank A. Sanders for his help with SEM imaging and acknowledge the use of the Precision Imaging Facility at NIST, Boulder for the SEM characterization reported in this work. This work was supported by the US Department of Energy, Office of Basic Energy Sciences, Division of Materials Sciences and Engineering under Award ER46921, contract DE-SC0010305 with the University of Colorado (B.S., A.M., T.L., I.I.S.), the Air Force Office of Scientific Research MURI Program, contract FA9550-12-1-0035 with the Rice University (J.M.T. and M.P.), the Air Force Office of Scientific Research FA9550-14-1-0111 (J.M.T.) and the Welch Foundation grant C-1668 (M.P.). B.S. also acknowledges support from the ICAM Branches Cost Sharing Fund and the Welch Foundation.

**Author contributions**

B.S., N.B., A.M., T.L. and D.E.T. performed experimental work. G.C. and J.M.T. provided graphene oxide materials. B.S., M.P. and I.I.S. analysed experimental results. I.I.S., M.P. and B.S. conceived and designed the project, and I.I.S. and B.S. directed the project. B.S. and I.I.S. wrote the manuscript with an input from all co-authors.

# Figures

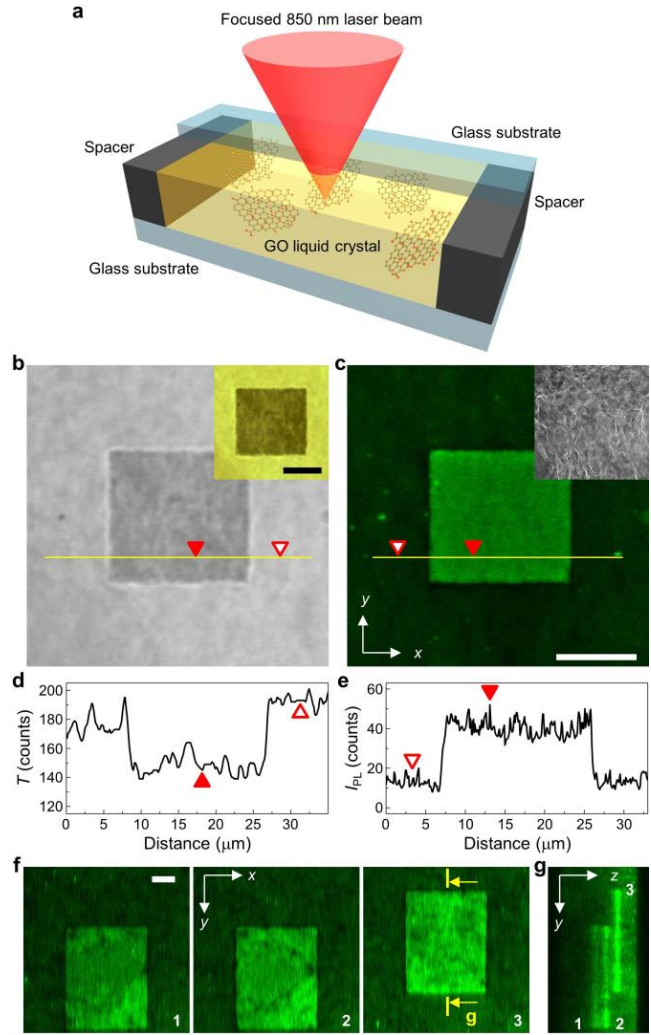

**Figure 1 | Laser-induced reduction of GO flakes in aqueous dispersions.** (a) Schematic diagram of the experiment. (b) Transmission-mode optical micrograph of an rGO plane (a dark square) in the aqueous dispersion of GO flakes obtained using 850 nm laser light; the inset shows the same sample area obtained using visible white light. (c) Photoluminescence texture of the rGO plane (a bright square) shown in b; the inset ($6 \times 6$ μm$^2$) shows an SEM image of a surface of the solid rGO plane microparticle retrieved from a discotic nematic LC sample. (d) Transmission and (e) photoluminescence along the yellow line, respectively, in b and c; magenta filled and empty triangles show intensities in corresponding points marked in b and c. (f) In-plane *x-y* and (g) cross-section *z-y* photoluminescent textures of three rGO planes produced in a thick ($d \approx 33$ μm) sample of the aqueous dispersion of GO flakes; a vertical cross-section z-y texture in g was obtained along the line marked by yellow arrows and a letter g in f. Scale bars, 10 μm.

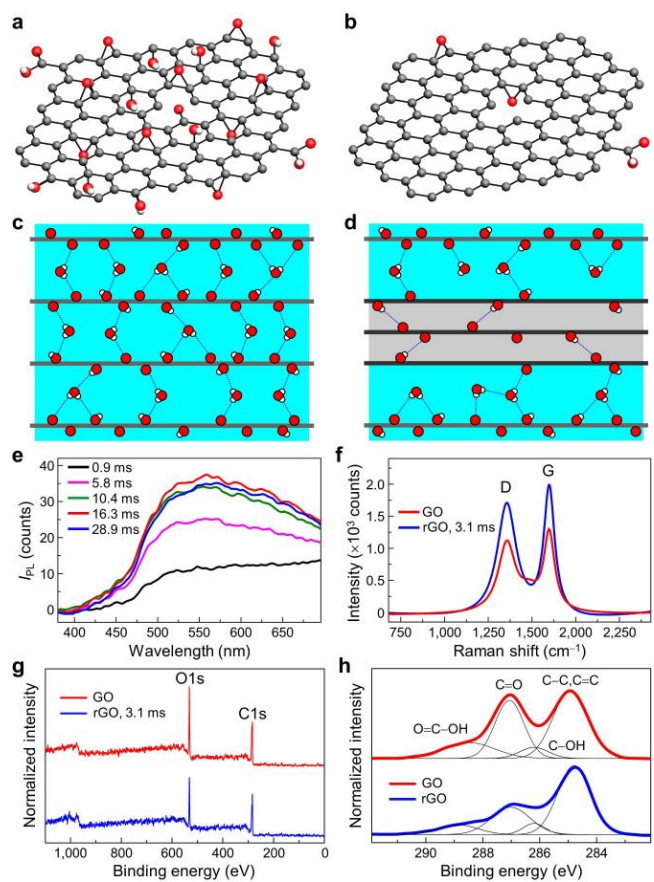

**Figure 2 | Laser-induced reduction of GO flakes in aqueous dispersions and their characterization.** (a,b) Simplified structure of (a) pristine and (b) reduced GO flakes. Dark grey, red and light grey spheres depict carbon, oxygen and hydrogen atoms, respectively. (c,d) Schematic diagram of (c) hydrophilic GO and (d) hydrophobic rGO flakes (thick grey lines) in a water (blue colour) solution; the space between rGO flakes (grey colour) is free of water. (e) Spectral dependence of photoluminescence intensity $I_{PL}$ at different dwell times and $E \approx 61$ mJ cm$^{-2}$. (f) Raman spectra of pristine and laser beam treated (at laser fluence of $E \approx 69$ mJ cm$^{-2}$) GO flakes near D and G peaks. (g) Survey XPS spectra of GO and rGO areas within the same sample. (h) C1s XPS spectra of GO and rGO.

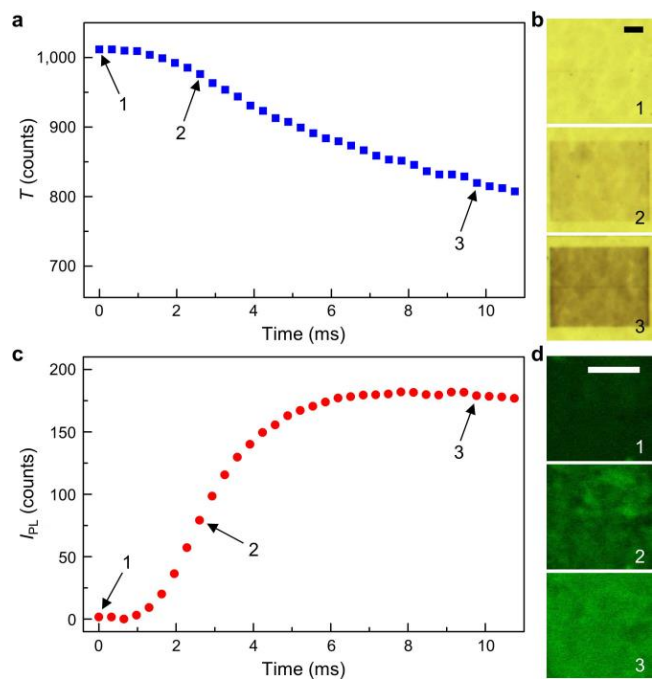

**Figure 3 | Optical properties of rGO microstructures suspended within aqueous GO dispersions.**
(a) Intensity of laser light at 850 nm transmitted through a treated area in a thick ($d \approx 27$ μm) sample versus dwell time. (b) Optical microscopy textures of sample areas with rGO structures after different irradiation times. Numbers marked on the textures correspond to points indicated by arrows in a. (c) Photoluminescence intensity of rGO depending on the dwell time of irradiation. (d) Nonlinear microscopy in-plane textures of sample areas irradiated for different times. Numbers marked on the textures correspond to points indicated by arrows in c. The used laser fluence was $E \approx 61$ mJ cm$^{-2}$. Scale bars, 10 μm.

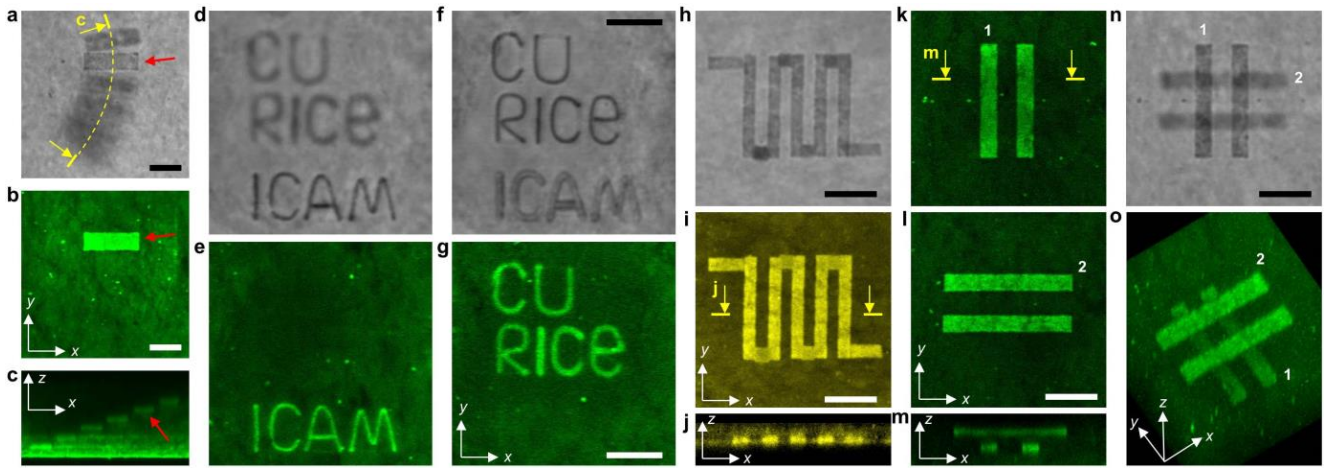

**Figure 4 | Micropatterning of rGO structures in a bulk of a GO flakes aqueous dispersion.** (a) Optical transmission and (b) photoluminescence in-plane textures of a structure of curved steps. (c) Photoluminescence cross-sectional texture of stair-like structure acquired along a dashed yellow line marked by yellow arrows and a letter c in the panel a. Magenta arrows in b and c point to the same step. (d,f) Optical transmission and (e,g) photoluminescence micrographs of different parts of the text patterned at different depths. (h) Optical transmission and (i,j) photoluminescence textures of a micropatterned resistor; a vertical cross-section $z$-$x$ texture in j was obtained along the line marked by yellow arrows and a letter j in i. (k–m) Photoluminescence and (n) optical transmission micrographs of a micropattern mimicking a thin film capacitor or a field transistor; a cross-section $z$-$x$ texture in m was obtained along the line marked by yellow arrows and a letter m in k. The same rGO buses/stripes in k–n are noted by numbers 1 and 2. (o) Three-dimensional perspective view of the structure in k–n reconstructed from photoluminescence data using Fluoview. Scale bars, 10 μm.

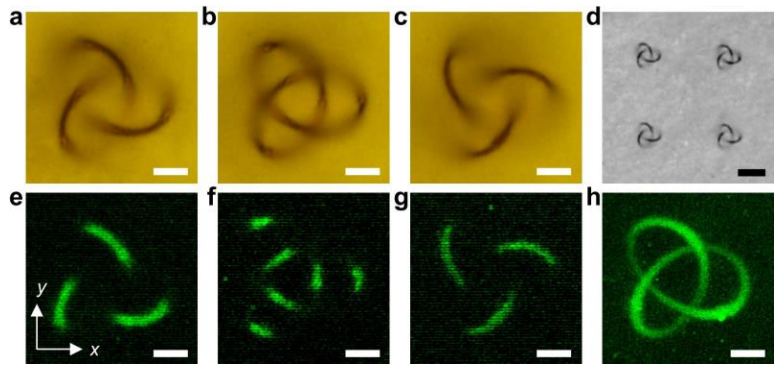

**Figure 5 | Topologically nontrivial rGO structures in a bulk of an aqueous GO flakes dispersion.** (a–c) Optical bright-field microscopy textures of an rGO left-handed trefoil knot focused on the (a) bottom, (b) middle and (c) top of the knot. (d) Optical bright-field micrograph of an array of rGO trefoil knots. (e–g) Photoluminescence textures of an rGO trefoil knot scanned in the (e) bottom, (f) middle and (g) top planes of the knot. (h) Three-dimensional perspective view of an rGO left-handed knot reconstructed from photoluminescence data using Fluoview. Reduction of the knot structure was done at $E \approx 30$ mJ cm$^{-2}$ and a laser beam scanning speed of 37 μm s$^{-1}$. Scale bars, 20 μm black colour and 5 μm white colour.

# Supplementary Figures

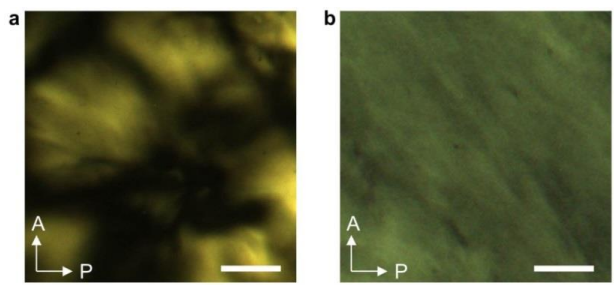

**Supplementary Figure S1 | Nematic liquid crystal of GO flakes dispersed in water.** Polarizing microscopy images of a (**a**) multidomain Schlieren and (**b**) homogeneous texture of a GO nematic liquid crystal at concentration 0.7 wt% in a cell of thickness ~ 25 µm. Scale bar, 5 µm.

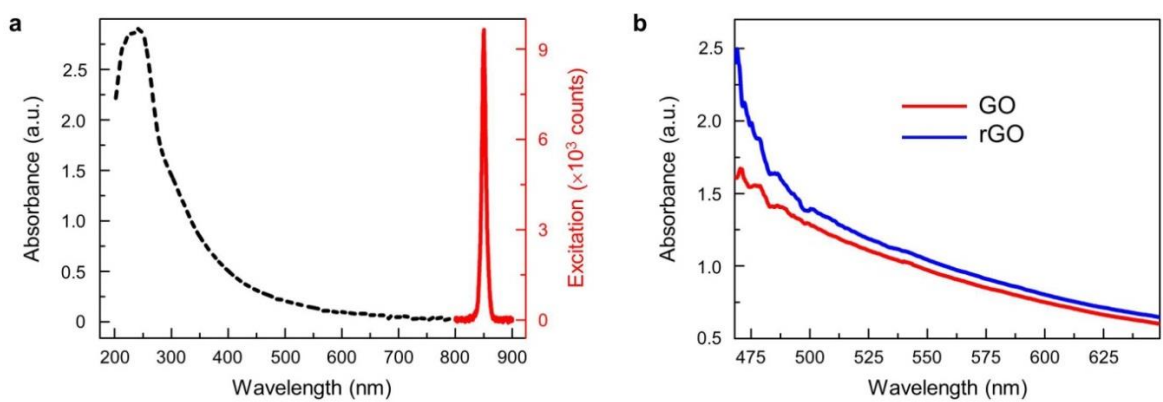

**Supplementary Figure S2 | Optical properties of GO flakes dispersed in water.** (**a**) Absorbance of aqueous GO flakes in a liquid crystal phase at 0.25 wt% (dashed black line) and a profile of an excitation laser beam with a central wavelength at 850 nm (solid red line). (**b**) Absorbance of the GO and rGO areas in a 10 µm thick liquid crystal sample at 0.8 wt%; the GO flakes were reduced to rGO at a fluence $E \approx 81$ mJ cm$^{-2}$ and a dwell time of 3 ms.

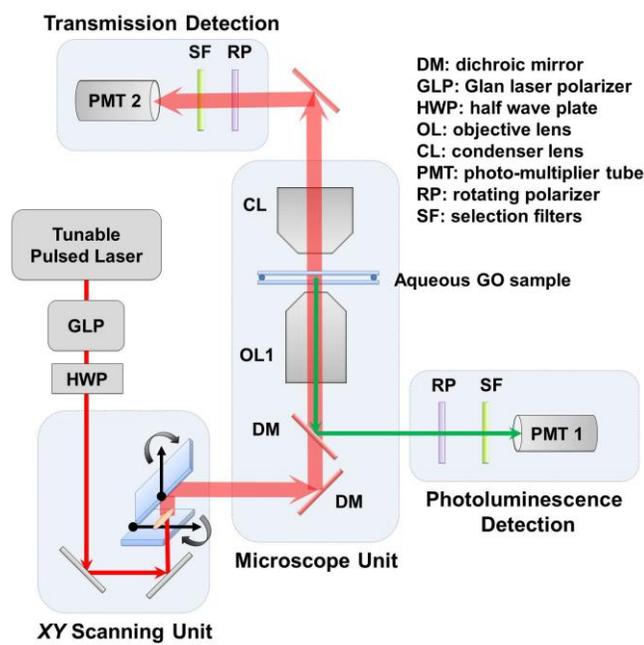

**Supplementary Figure S3 | Laser-induced reduction and imaging setup.** A schematic diagram of a nonlinear optical imaging setup based on a tunable pulsed laser, *xy* scanning mirror, and inverted microscope with multiple detection channels. Red lines show an excitation laser beam at 850 nm and green lines show photoluminescence signal collected using backward detection.

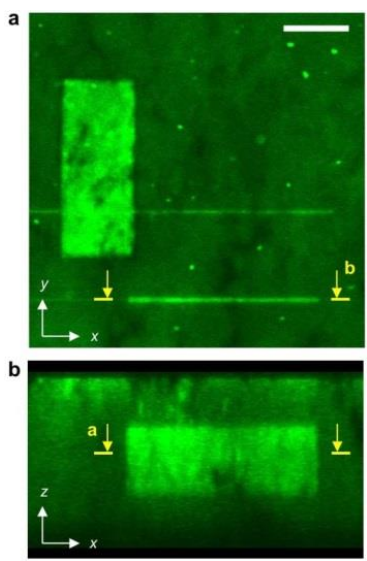

**Supplementary Figure S4 | Three-dimensional reduction of GO flakes in the bulk of a liquid crystal sample.** Photoluminescence imaging within (**a**) in-plane and (**b**) vertical cross-sectional views of the rGO structure reduced in the bulk of a liquid crystal of aqueous GO flakes. Scale bar, 10 μm.

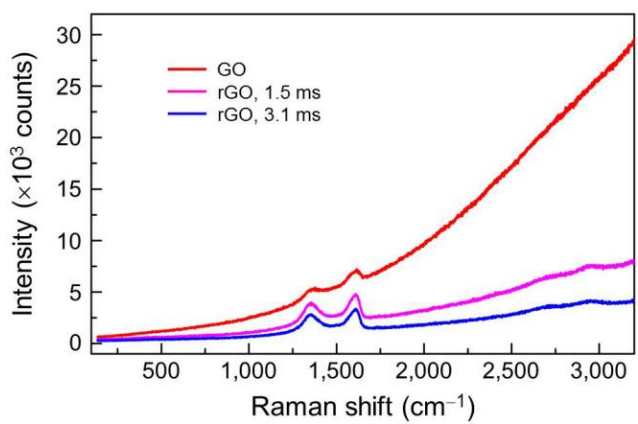

**Supplementary Figure S5 | Raman spectra of aqueous GO and rGO.** Raman spectra of aqueous GO and rGO flakes in a water solution before and after laser-induced reduction at laser fluence $E \approx 69$ mJ cm$^{-2}$.

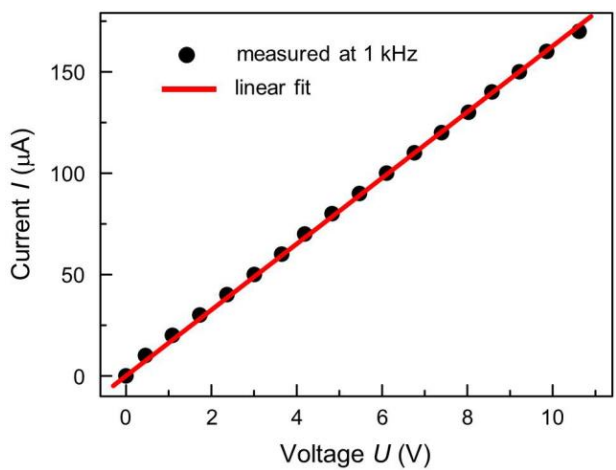

**Supplementary Figure S6 | Current-voltage dependence of an rGO thin film resistor.** Current-voltage characteristics of a single rGO ribbon of dimensions ~600 μm × 50 μm × 0.5 μm (length × width × thickness) that was micropatterned using laser reduction. The resistance data for the thin film of rGO were measured by a multimeter at the frequency of 1 kHz. Resistance of the rGO-resistor was evaluated from a linear fit as $R \approx 61.5$ kΩ.

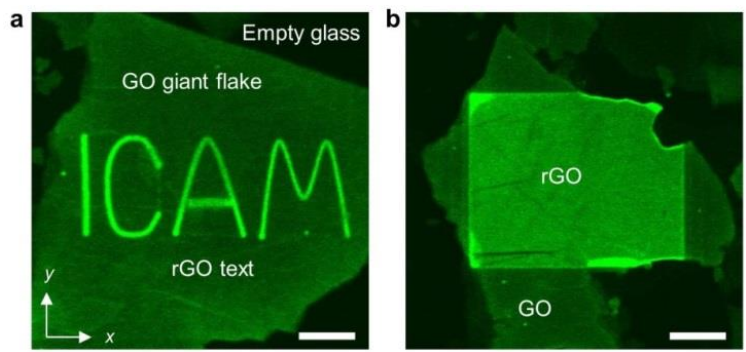

**Supplementary Figure S7 | Micropatterns of rGO in a dried thin film of GO flakes.** (**a**) Photoluminescence texture of rGO text pattern displaying "ICAM" (one of the funding agencies that supported the research) on a single giant GO flake deposited on a glass surface. (**b**) Photoluminescence texture of rGO square-like pattern in the GO dried thin film; rGO obtained by reduction at a laser fluence $E \approx 61$ mJ cm$^{-2}$ and a dwell time ~ 1 ms. Scale bar, 5 μm.

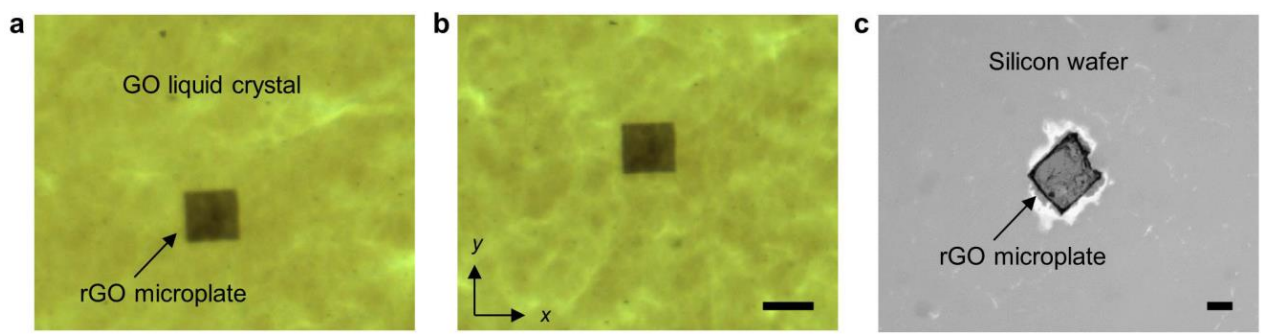

**Supplementary Figure S8 | Mechanical stability of rGO microstructures.** (**a,b**) Optical transmission-mode bright field micrographs of a laser-reduced colloidal rGO square plate moving as a whole piece from the initial position in **a** to a position in **b** under the influence of a shearing flow in a GO liquid crystal sample. (**c**) An optical reflective microscopy texture of a rectangular rGO microplate on a surface of a silicon wafer retrieved from a GO liquid crystal sample after laser-induced reduction and micropatterning. Bright irregular pattern around the rGO microplate shows a residue due to washing the rGO microparticle in water and drying on a silicon wafer. Scale bars, 15 μm.

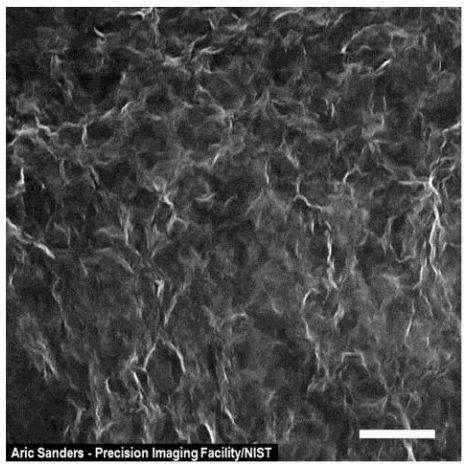

**Supplementary Figure S9 | Scanning electron microscopy image showing surface morphology of rGO microparticles.** The rGO microparticle corresponds to that shown in Supplementary Fig. 8c and was retrieved from an aqueous GO sample after laser-induced reduction. Scale bar, 1 μm.

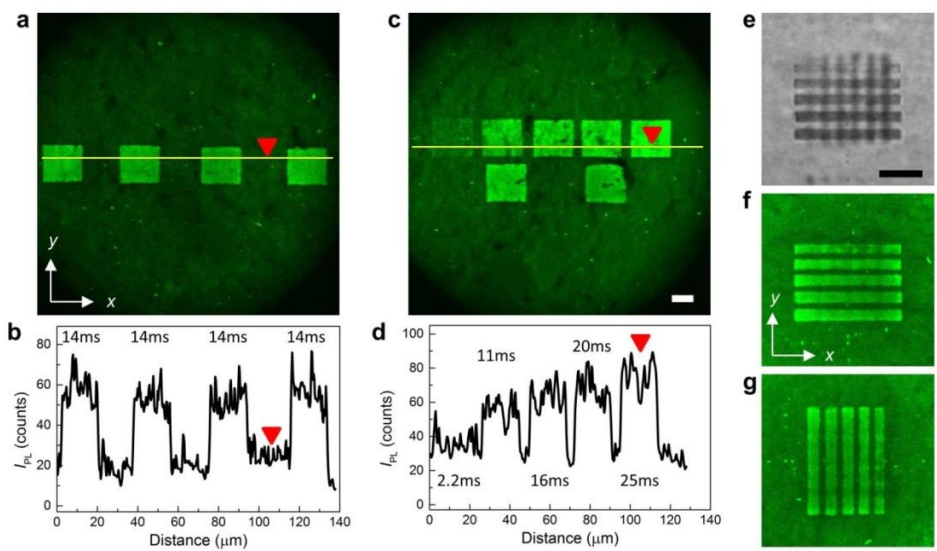

**Supplementary Figure S10 | Photoluminescence properties and micropatterns of rGO.** (**a,c**) Photoluminescence texture of rGO areas (bright square) in the aqueous dispersion of GO flakes obtained by laser-induced reduction using pulsed 850 nm laser beam at $E \approx 61$ mJ cm$^{-2}$ and different dwell times. (**b,d**) Photoluminescence intensity along the yellow line respectively in **a** and **c** depending on the dwell time. (**e**) Optical and (**f, g**) photoluminescence micrographs of perpendicular rGO stripes reduced on different depth levels in a cell. Scale bars, 10 μm.

**Supplementary Movie 1**: Three-dimensional photoluminescence imaging using 850 nm excitation light, revealing laser-patterned multi-level rGO microstructures in the form of text suspended within the aqueous GO flakes forming liquid crystal sample. The movie shows the sequence of two-dimensional scan through the thickness of the sample.

**Supplementary Movie 2**: Three-dimensional photoluminescence imaging using 850 nm excitation light, revealing laser-patterned structures that can be envisaged as prototypes of thin film capacitors or field transistors.

**Supplementary Movie 3**: A movie composed of a series of different three-dimensional perspective views of a colloidal particle in the form of a rGO-based trefoil knot reconstructed from the data of photoluminescence imaging performed using at 850 nm.